\documentclass[a4paper,prd,superscriptaddress,onecolumn,floatfix]{revtex4}

\usepackage[latin1]{inputenc}
\usepackage{amsfonts,amssymb,amsmath}
\usepackage{graphicx,epstopdf}

\usepackage[colorlinks=true, citecolor=blue, urlcolor=blue, linkcolor=red, bookmarks=true]{hyperref}
\usepackage{orcidlink}
\begin{document}

\renewcommand{\arraystretch}{1.3}
\setlength{\tabcolsep}{1ex}

\title{Generating exact polytropes in non-conservative unimodular geometries}

\author{Sudan Hansraj\orcidlink{0000-0002-8305-7015}}%
\email[Email: ]{hansrajs@ukzn.ac.za} 
\affiliation{Astrophysics Research Centre, School of Mathematics, Statistics and Computer Science, University of KwaZulu-Natal, Private Bag X54001, Durban 4000, South Africa}

\author{Chevarra Hansraj\orcidlink{0000-0001-5304-7433}}%
\email[Email: ]{chevarrahansraj@gmail.com} 
\affiliation{Astrophysics Research Centre, School of Mathematics, Statistics and Computer Science, University of KwaZulu-Natal, Private Bag X54001, Durban 4000, South Africa}

\author{Njabulo Mkhize\orcidlink{0000-0001-6893-6305}}%
\email[Email: ]{MkhizeN@unizulu.ac.za} 
\affiliation{Astrophysics Research Centre, School of Mathematics, Statistics and Computer Science, University of KwaZulu-Natal, Private Bag X54001, Durban 4000, South Africa}

\author{Abdelghani Errehymy\orcidlink{0000-0002-0253-3578}}%
\email[Email: ]{abdelghani.errehymy@gmail.com}
\affiliation{Astrophysics Research Centre, School of Mathematics, Statistics and Computer Science, University of KwaZulu-Natal, Private Bag X54001, Durban 4000, South Africa}

\author{Christian G. B\"{o}hmer\orcidlink{0000-0002-9066-5967}} %
\email[Email: ]{c.boehmer@ucl.ac.uk} 
\affiliation{Department of Mathematics, University College London, \\ Gower Street, London WC1E 6BT, UK}
\affiliation{Astrophysics Research Centre, School of Mathematics, Statistics and Computer Science, University of KwaZulu-Natal, Private Bag X54001, Durban 4000, South Africa}

\date{\today}

\begin{abstract}
The trace-free Einstein equations contain one equation less than the complete field equations. In a static and spherically symmetric spacetime, the number of field equations is thus reduced to two. The equation of pressure isotropy of general relativity, however, is preserved thus showing that any known perfect fluid spacetime is a suitable candidate for the trace-free scenario. The extra freedom in imposing two constraints may now be exploited to include polytopes, something that is difficult in general relativity. The point here is that using any known exact solution one can find a polytropic star for various values of the polytropic index. One arrives at Tolman-Oppenheimer-Volkoff type equations and can study their solutions explicitly. Two examples of well-known stellar distributions that generate polytropes with physically reasonable behaviour are discussed. These models are regular, exhibit a sound speed that is never superluminal and are adiabatically stable in the sense of Chandrasekhar. We investigate a compactness measure confirming that our results are consistent with some observational data.
\end{abstract}

\maketitle

\section{Introduction}
\label{intro:Sec}

When studying compact objects like neutron stars, brown dwarfs, white dwarfs, and also some main-sequence stars one can often model the matter content of such objects using a polytropic equation of state (EoS). Any EoS expresses mathematically the relationships between the thermodynamical properties of the state of matter such as the pressure $p$, density $\rho$ and temperature $T$. In Newtonian and relativistic astrophysics one generally works with a relationship of the form $p = p(\rho)$, or equivalently $\rho=\rho(p)$. The knowledge of the EoS is vital to understand the internal stellar structure which is helpful when studying the energy transport mechanisms and conditions needed for certain nuclear reactions to occur~\cite{Lattimer:2000nx,baturin:2003nx}. Specifically certain ultrahigh density stars are laboratories for nucleosynthesis~\cite{Battistelli:2022fco}. The EoS also assists in tracing the evolution of stars as they age and their nuclear fuel is exhausted until they reach their final fate which could be, for example, a white dwarf, a neutron star or a black hole~\cite{daSilvaSchneider:2020ddu,Lyra:2022qmg,Holgado:2021vaq}. Moreover, stars tend to achieve equilibrium and the EoS is helpful in analysing the stability of the star and the conditions under which instabilities may arise causing gravitational collapse or expansion~\cite{Joshi:2011rlc,Joshi:2011zm}. In addition, the EoS also conveys information about the observable properties of stars such as luminosity, surface temperature, spectral features and internal composition. In turn this assists in ruling out certain modified theories of gravity since they may not comport with the observational evidence. In fact recent gravitational wave detections from the binary neutron star merger GW170817 have introduced stringent limits on the masses and radii of coalescing stars~\cite{Lattimer:2020tot,LIGOScientific:2017vwq}. 

Constructing viable models of stellar distributions requires solving Einstein's field equations with a physically reasonable source term -- most often a perfect fluid is a reasonable starting point. In classical general relativity (GR) an exact solution is regarded the most desirable form of solution, compared to a numerical solution. Typically in GR, the prescription of an EoS closes the system of field equations, however, this does not make them easy to solve. Quite contrary, to date, no exact solution for realistic isotropic stars have emerged, even by considering the simplest equation of state, that of a linear barotropic equation of state. Consequently, no polytropic solutions describing realistic stars are known. Numerical treatments of both these configurations are well known~\cite{Nilsson:2000zf,Nilsson:2000zg}. The only successful attempt reported in the literature for a perfect fluid with a linear equation of state is due to Saslaw {\it{et al}}~\cite{Saslaw:1996bs}. However, aforesaid authors overdetermined the system of field equations by requiring an inverse square fall-off of both pressure and density. Fortuitously, the solution they found is actually correct and all the field equations are satisfied. However, the difficulty of the Saslaw model is that it does not exhibit a boundary consequently, it may only be used to model a universe filled with perfect fluid. It does not apply to stars. The simple equation of state $p = \gamma \rho$ when introduced into the Einstein field equations, results in a  master equation that is intractable and perhaps impossible to solve. 

Historically, nearly all exact solutions for spherically symmetric isotropic matter have been found through mathematical assumptions made to solve the complicated system of nonlinear equations. This is possible because the under-determined system of field equations has one free function. This choice is sufficient to successfully generate many solutions, over 120 exact solutions reported here~\cite{Delgaty:1998uy} for instance. The well-known eight Tolman metrics~\cite{Tolman:1939jz}, found as early as 1939, were obtained by arranging the variables in a master equation so that certain nonlinearities can be made to vanish. The caveat with this approach is that the physical properties of these solutions, such as the resulting EoS, generally have no strong physical interpretation, see also~\cite{Fodor:2000gu}.

Throughout this work we will consider polytropic equations of state which are given by
\begin{equation}
    p = \gamma\, \rho^{\Gamma}, \qquad \Gamma = 1+\frac{1}{n},
\end{equation}
where $\gamma$ is a positive constant and $n$ or $\Gamma$ is called the polytropic index, depending slightly on conventions. The standard approach  is to start with the continuity equation and by defining the mass function. In the Newtonian setting, this leads to the Lane-Emden equation~\cite{Santana:2022vmw,Saad:2021dwp}. The first to investigate the relativistic polytrope for static compressible fluid spheres was Tooper~\cite{tooper} who went on to solve the relativistic Lane-Emden equation exactly for $n =0$ and numerically for $n = 1, 0.5, 3$. In the Newtonian setting, the Lane-Emden equation integrates exactly for $n = 0, 1, 5$. However, this approach does not necessarily yields a closed form solution of the metric~\cite{Lindblom:2013kra}. Hence, the geometry for the stellar system can remain unknown. Different numerical values of $n$ describe different physical situations. 

For example, $n=0$ is an incompressible star with constant density, this is, the Schwarzschild interior solution. Neutron stars are represented by values $0.5 < n < 1$~\cite{Lattimer:2006xb,Flanagan:2007ix}. An index $n = 1.5$ approximates red giants, brown dwarfs or giant gaseous planets (like Jupiter)~\cite{chandra,hansen}. White dwarfs of low mass  can be modelled by $n = 1.5$  while a value of $n = 3$ is used for white dwarfs of higher masses~\cite{Sagert:2005fw}. The value $n=3$ is also used for some main-sequence stars, like our Sun. If $n=5$, in the Newtonian case, one finds a solution of infinite radius which does not model a compact object but could be used to model a stellar system.

An option worth exploring is the trace-free version of Einstein's equations which Einstein himself proposed in an effort to solve some issues with his equations. Also known as unimodular gravity, essentially what occurs is that the determinant of the metric tensor is set to unity. The net effect is that one field equation is lost and the conservation equation no longer holds. In the case of spherical symmetry, the number of independent field equations drops to 2 in four unknown functions whereas they were 3 equations in four unknowns in GR. energy-momentum conservation may now be reintroduced by hand, thus restoring the field equations back to their standard form. There are effectively no gains with this process and Visser~\cite{Visser:2003ge} has correctly argued that the trace-free equations are equivalent to Einstein's equations. Of course, the difference is the absence of energy conservation results in new physics and several authors discussed that the equivalence is only geometric~\cite{Darabi:2017coc,Hansraj:2020clg}. In fact, Ellis {\it{et al}}~\cite{Ellis:2010uc} argued that in the context of the trace-free equations, the cosmological constant is merely an integration constant and thus the problem of the inconsistency in the value of the cosmological constant from quantum field theory and observation, vanishes. The integration constant may now be at the scale of the thermodynamical variables and may even influence stellar structure as shown in~\cite{Hansraj:2017len} whereas the normal cosmological constant is too small to have any impact on astrophysical processes. Josset {\it {et al}} argued that a violation of energy conservation could account for the presence and effects of dark energy~\cite{josset}, also in the context of unimodular gravity.

The idea of dismissing energy conservation is not novel and can be included in a general framework of studying modified theories of gravity, see for example~\cite{Boehmer:2021aji,Boehmer:2023fyl}. The physical meaning of abandoning energy-momentum conservation is that the matter and geometry are coupled in a non-minimal way unlike in general relativity where the coupling is minimal. For example, see the work of Moradpour {\it {et al}}~\cite{moradpour}, where the aim was to study the proposal of a 4-index theory of gravity of Moulin~\cite{moulin} that contained general relativity when reduced to the usual 2-index version. 

About fifty years ago, Rastall~\cite{Rastall:1972swe,Rastall:1976uh} argued that setting the covariant divergence of the Einstein tensor to zero permits the energy-momentum tensor to have a divergence proportional to the gradient of the Ricci scalar. Rastall theory generates an isotropy equation that is identical to the Einstein equation which means that all solutions of Einstein gravity are solutions of Rastall gravity. The deviation occurs due to the violation of energy conservation. Visser~\cite{Visser:2017gpz} paradoxically acknowledged a violation of energy conservation in Rastall's proposal but still concluded that the theory was trivially equivalent to Einstein's theory. This equivalence is only in terms of geometry and not physics. Despite the violation of energy-momentum conservation, Rastall gravity may not be dismissed until experimental evidence arises that rules it out. For now, note the successes of Rastall theory: It is consistent with the age of the universe problem and the Hubble parameter~\cite{alrawaf1}, it comports with helium nucleosynthesis~\cite{alrawaf2}, it behaves as expected during gravitational lensing~\cite{abdel-rahman} and it accounts for the cosmic accelerated expansion of the universe where general relativity fails without invoking the mysterious dark matter for which no experimental evidence exists to date~\cite{alrawaf3,moradpour2}. Note that these physically well behaved phenomena are in spite of the violation of energy-momentum conservation. In fact another weakness of Rastall theory is the absence of a suitable Lagrangian however, this has not halted intensive investigations of the theory. It must also be noted that energy conservation violation is also evident in $f(R, T)$ gravity~\cite{Harko:2011kv} which has also been thoroughly studied in recent times.  In this formulation, the Lagrangian density is composed of the Ricci scalar $R$ as well as a term $T$ which is the trace of the energy-momentum tensor $T_{ab}$. Another theory that abandons energy-momentum conservation is Weyl-squared gravity~\cite{Kiefer:2017nmo}. 

The work is organised as follows: In Section~\ref{Sec2}, we recall key features of trace-free gravity and then derive the field equations in the following Section~\ref{Sec3}. There are 2 independent equations in 4 unknowns. In Section~\ref{Sec4}, we confirm that the Schwarzschild exterior metric is still the vacuum solution and Birkhoff's theorem is unaffected for trace-free gravity. We present two physically interesting cases of the Finch-Skea and Vaidya-Tikekar metrics and obtain exact models with good astrophysical properties using a polytropic equation of state in Section~\ref{Sec5}. The article ends with a discussion in Section~\ref{Sec6}.

\section{Trace free Einstein gravity}
\label{Sec2}

A comprehensive discussion of trace-free Einstein gravity and its relationship with unimodular gravity~\cite{Anderson:1971pn,Finkelstein:2000pg,Smolin:2009ti} is given in~\cite{Ellis:2010uc} and~\cite{Ellis:2013uxa}. Nevertheless, we recollect a few basic points. In standard general relativity (GR) the gravitational field is governed by the Einstein field equations
\begin{equation}
    G_{ab}:= R_{ab} - \frac{1}{2} R g_{ab} = \kappa\, T_{ab},
    \label{EFE}
\end{equation}
where we set the gravitational coupling $\kappa = 8\pi G/c^4$ to unity. Also we use geometrized units throughout the paper and switch to SI units when discussing the physical properties. The conservation equations follow naturally
\begin{equation}
    \label{cons}
    \nabla_b G^{ab}=0 \quad \Rightarrow \quad \nabla_b T^{ab}=0.
\end{equation} 
Denoting by a hat the trace-free part of a symmetric tensor, we may write
\begin{equation}
    \hat{G}_{ab} = R_{ab} - \frac{1}{4}Rg_{ab}, \qquad 
    \hat{T}_{ab} = T_{ab} -\frac{1}{4} Tg_{ab}, \quad \Rightarrow \quad 
    \hat{G}^a_a = 0, \quad \hat{T}^a_a=0,
\end{equation}
then~(\ref{EFE}) implies
\begin{equation}
\label{TFE}
    \hat{G}_{ab} = \hat{T}_{ab}, \quad \Leftrightarrow \quad 
    R_{ab} -\frac{1}{4}Rg_{ab} = T_{ab} - \frac{1}{4} Tg_{ab},
\end{equation}
which are the trace-free Einstein field equations (TFE). These are now the equations of motion we use for the gravitational field. Observe that the conservation laws 
\begin{equation}
    \label{cons2}
    \nabla_b T^{ab} = 0,
\end{equation} 
no longer follow as a natural consequence of the field equations. They could be inserted to the system as an additional constraint. Then taking the divergence of~(\ref{TFE}) and integrating gives
\begin{equation}
    G_{ab} + \Lambda g_{ab} = T_{ab},
    \label{EFE1}
\end{equation}
where the  constant $\Lambda$ usually called the cosmological constant is now  a mere constant of integration and has no connection with vacuum energy~\cite{Weinberg:1988cp,Ellis:2010uc}. In GR the cosmological constant has a very small value in order to be compatible with solar system observations. It is too small to impact the structure of stars and is consequently ignored in astrophysical modelling. However, from a mathematical point of view, the Schwarzschid interior solution can be extended to inlude the cosmological constant which leads to new classes of solutions~\cite{Boehmer:2003uz,Boehmer:2004nu,Balaguera-Antolinez:2004ytv,Boehmer:2005kk} with interesting properties.

In the framework of trace-free gravity the constant $\Lambda$ may be at the scale of the other variables and is not necessarily negligible. Thus TFE solutions are the same as the GR solutions with an arbitrary cosmological constant (which can even be zero) when energy conservation is added. If energy conservation is absent then the equations~(\ref{TFE}) have different physical consequences. We are interested in this latter case. 

\section{TFE Field Equations}
\label{Sec3}

\subsection{Spherically symmetric field equations}

The static spherically symmetric spacetime in coordinates $(t, r, \theta, \phi)$ is
\begin{equation}
    ds^2 = -e^{2\nu(r)}dt^2  + e^{2\lambda (r)} dr^2 + r^2 \left(d\theta^2 +\sin^2 \theta d\phi^2\right), 
    \label{1}
\end{equation}
where the metric functions $\nu$ and $\lambda$ are functions of the radial coordinate $r$ only. The fluid's 4-velocity is $ u^a = e^{-\nu} \delta^a_0$ and we consider a perfect fluid source with energy-momentum tensor $T_{ab} = (\rho + p)u_a u_b + p g_{ab}$.

The trace free Einstein tensor are
\begin{eqnarray}
    \hat{G}_{tt} &=& \frac{e^{2(\nu - \lambda)} }{2r^2} \left(r^2(\nu''+\nu'^2 - \nu'\lambda') +2r(\nu' + \lambda') + e^{2\lambda} - 1 \right),  
    \label{2a} \\[1ex]
    \hat{G}_{rr} &=& \frac{1}{2r^2} \left(2r(\nu' + \lambda') -r^2(\nu'' + \nu'^2 -\nu'\lambda') -e^{2\lambda} + 1\right),  
    \label{2b} \\[1ex]
    \hat{G}_{\theta \theta} &=& \frac{e^{-2\lambda}}{2}\left(r^2(\nu'' + \nu'^2 - \lambda'\nu') + e^{2\lambda} -1 \right),
    \label{2c}
\end{eqnarray}
and $\hat{G}_{\phi \phi} = \sin^2\negmedspace\theta\, \hat{G}_{\theta \theta}$. The trace-free components of the energy-momentum tensor are
\begin{equation}
    \hat{T}_{ab} = \left( \frac{3}{4}(\rho + p)e^{2\nu}, \frac{1}{4}(\rho + p)e^{2\lambda}, \frac{r^2}{4} (\rho + p), \frac{r^2 \sin^2 \theta}{4} (\rho + p)\right).
    \label{3}
\end{equation}
Ordinarily, in the Einstein field equations the $T_{00}$ component is independent of pressure but that is not the case in here. Consequently, studying the the constant density case in trace-free gravity is much more nontrivial when energy conservation is abandoned. 

The TFE field equations now have the form
\begin{eqnarray}
    (\nu'' + \nu'^2   - \nu'\lambda' ) +\frac{2}{r}\left(\nu' + \lambda'\right) +\frac{e^{2\lambda}-1}{r^2}&=&\frac{3}{2}(\rho + p)e^{2\lambda},
    \label{4a} \\[1ex]
    \frac{2}{r}\left(\nu' + \lambda'\right)  -(\nu''  +\nu'^2 - \nu' \lambda') -\frac{e^{2\lambda} -1}{r^2} &=& \frac{1}{2}(\rho + p)e^{2\lambda},
    \label{4b} \\[1ex]
    (\nu'' + \nu'^2 - \nu'\lambda') + \frac{e^{2\lambda}-1}{r^2}&=& \frac{1}{2}(\rho + p)e^{2\lambda}. 
    \label{4c}
\end{eqnarray}
Note that the trace-free nature of the field equations shows that only the combination $\rho+p$ enters the field equations. It is the equation of state that will allow us to separate these two quantities. It is easy to verify that these three equations simultaneously imply the master equation
\begin{equation}
    r^2(\nu'' + \nu'^2 - \nu'\lambda') -r(\nu'+\lambda') +(e^{2\lambda}-1) = 0, 
    \label{4d}
\end{equation}
which is also taken as the equation of pressure isotropy. We have the further equation
\begin{equation}
    \frac{2}{r}\left(\nu' + \lambda'\right) = (\rho + p)e^{2\lambda}.
    \label{4d1}
\end{equation}
The two equations~(\ref{4d}) and~(\ref{4d1}) imply all three of~(\ref{4a})--(\ref{4c}). These are the stellar structure equations are now two equations in four.

The transformations $x=Cr^2$ ($C>0$ a constant), $e^{\nu(r)} = y(x)$ and $e^{-2\lambda (r)} = Z(x)$ are known to convert the master isotropy equation~(\ref{4d}) to a linear second order ordinary differential equation in $y$. This is a significant insight into removing nonlinearities. These substitutions originated  by Buchdahl~\cite{buchdahl} also reduce (\ref{4d}) to a first order linear equation in $Z$ which in turn allows for the separation of the variables. Unfortunately the final forms are too complicated for practically detecting new exact solutions. With these transformations~(\ref{4d}) reduces to
\begin{equation}
    4x^2 Z\ddot{y} + 2x^2\dot{Z}\dot{y} + (\dot{Z}x - Z +1) y = 0,
    \label{4f}
\end{equation}
where the dots denote derivatives with respect to $x$. This is exactly the same pressure isotropy equation as in standard Einstein gravity~\cite{Chilambwe:2015fia}. On the other hand, Eq.~(\ref{4d1}) assumes the form
\begin{equation}
    4Z\dot{y}-2\dot{Z}y=\frac{\rho+p}{C}y.
    \label{4f1}
\end{equation}
There are already over 120 known exact solutions of equation~(\ref{4f}) published in the literature. We can therefore utilise these as a starting point. It makes sense to use solutions that have been shown to be physically viable in Einstein theory although we cannot rule out other metrics satisfying~(\ref{4f}) since the dynamics of the present problem are quite different from the Einstein situation. For example, one can speculate on a particular form for $\rho + p$ and then attempt to find a simultaneous solution of~(\ref{4f}) and~(\ref{4f1}). This would be a formidable project, barring a few simple cases, with slim chances of success in general. As remarked above there are 2 gravitational field equations in 4 unknowns. So two choices are left open.

\subsection{Vacuum solution}
\label{Sec4}
 
In GR Birkhoff's theorem states that a spherically symmetric solution of the vacuum field equations is static and consequently, the exterior Schwarzschild metric is the unique vacuum solution. The situation turns out to be the same in trace-free gravity which we confirm below. The vacuum metric is determined equations~(\ref{4a})--(\ref{4c}) with $\rho = p = 0$, for the exterior. Then, as in Einstein gravity, the condition $\nu'=-\lambda'$ emerges and there is only one independent equation 
\begin{equation}
    (\nu'' + \nu'^2 - \nu'\lambda') + \frac{e^{2\lambda}-1}{r^2} = 0,
    \label{5a}
\end{equation}
for example~(\ref{4c}) to be solved. Putting $\nu' = -\lambda'$ into~(\ref{5a}) gives
\begin{equation}
    r^2(\lambda'' - 2\lambda'^2)  - (e^{2\lambda} -1) = 0,
    \label{5b}
\end{equation}
which is second order and nonlinear. Solving~(\ref{5b}) is made easier by invoking the substitution $e^{2\lambda (r) } = B(r)$. Then~(\ref{5b}) becomes
\begin{equation}
    r^2 (BB'' - 2B'^2)+2B^2(1-B) = 0,
    \label{5c}
\end{equation}
which is still nonlinear but solvable. The exact solution is given by 
\begin{equation}
    B(r) = \frac{r}{c_2 r^3+ r-c_1} = e^{2\lambda} = e^{-2\nu}.  
    \label{5d}
\end{equation}
This is identical to the Schwarzschild-de Sitter metric. Note that $c_1$ and $c_2$ are integration constants. The line element of the vacuum metric may be written as 
\begin{equation}
    ds^2 = -\left(1 -\frac{2M}{r} + \frac{\Lambda}{3} r^2 \right) dt^2 + 
    \left(1 -\frac{2M}{r} + \frac{\Lambda}{3} r^2 \right)^{-1} dr^2 + 
    r^2 (d\theta^2 + \sin^2\negmedspace\theta d\phi^2),
    \label{5e}
\end{equation}
where we have made the standard identification $c_1 = 2M$ as the active gravitational mass as measured by an observer at spatial infinity and $c_2 = \Lambda/3$  which resembles the cosmological constant. It is now simply an integration constant of arbitrary scale. The radius $r=R$ denotes the bounding radius of the fluid distribution. One defines this radius to be the vanishing pressure surface, so that $p(R)=0$. This metric is the Schwarzschild-de Sitter metric and it should be used when matching an interior spacetime with the exterior geometry.

\subsection{Physical conditions and polytropes}
\label{Sec5}

In order for a model to have physical validity the following elementary constraints are usually imposed. The energy density and pressure should be positive, the pressure must vanish at some finite radius which defines the boundary of the object. In the absence of a boundary one could consider such solutions as cosmological fluids or as models for entire galaxies, see for example~\cite{Saslaw:1996bs} where an isothermal universe is modelled. In the case of a stellar distribution the interior and exterior metrics should match across a boundary. The metric potentials should match and by the Israel-Darmois junction condition the continuity of the second fundamental forms is equivalent to the vanishing of the radial pressure~\cite{israel1,Israel:1966rt,darmois,Lake:2017wlx}. 

The expressions controlling the weak $\rho-p$, strong $\rho +p$ and dominant $\rho + 3p$ energy conditions should all be positive within the star. The sound speed squared in both the radial and tangential directions should be less than unity so that the fluid is never acausal, this means the speed of sound is bounded by the speed of light. This means we require
\begin{equation}
    0 < \frac{dp_r}{d\rho} < 1,\qquad 
    0< \frac{dp_t}{d\rho} <1.
\end{equation}
According to Chandrasekhar~\cite{Chandrasekhar:1964zz,Chandrasekhar:1964zza} the following should also hold
\begin{equation}
    \left(\frac{\rho+p_r}{p_r}\right) \frac{dp_r}{d\rho} > \frac{4}{3},\qquad
    \left(\frac{\rho+p_t}{p_t}\right) \frac{dp_t}{d\rho} > \frac{4}{3},
\end{equation}
to maintain the adiabatic stability of the fluid sphere. 

Any exact solution for the functions $(Z, y)$ determines the quantity $\rho + p$ in~(\ref{4f1}). This allows us to immediately check the strong energy condition. However, without an explicit equation of state one cannot isolate the energy density of pressure. As discussed in the above, when choosing, for example, a polytropic EoS one can now explicitly find expressions for the energy density and the pressure and study the physical properties of such solutions. 

\section{The Finch-Skea seed metric}

The metric due to Finch and Skea~\cite{finch} has been thoroughly investigated, when proposing this metric, it was checked that it was in agreement with previous astrophysical data~\cite{Walecka:1975ft}. We therefore elect to use it in our study. It is given by 
\begin{eqnarray}
    Z&=& \frac{1}{v^2}, \qquad v=\sqrt{1+x},
    \label{10a} \\
    y&=& \sqrt{\frac{2}{\pi }} \left(\sin v \left( A- B v\right)-\cos v \left( B+ A v\right)\right),
    \label{10b}
\end{eqnarray}
so that one can compute, using Eq.~(\ref{4f1})
\begin{equation}
    \rho + p = 
    C\, \frac{2 \left(\tan v \left(Bv-A(v^2 +1)\right)+ \left(A v
    + B (v^2+1)\right)\right)}
    {v^4 \left(\tan v \left(B v-A\right)+ \left(A v+B\right)\right)} := f(v).
    \label{11}
\end{equation}

\subsection{Linear barotropic equation of state}

We denote the right-hand side of~(\ref{11}) to be $f(v)$. Then, assuming the linear equation of state $p = \gamma \rho$, one immediately obtains
\begin{equation}
    \rho = \frac{1}{1+\gamma} f(v), \qquad 
    p = \frac{\gamma}{1+\gamma} f(v).
\end{equation}
Of course, the sound speed squared is given by $\gamma$ which means we will assume $0<\gamma <1$ to satisfy causality.

The various energy conditions take the simple forms
\begin{eqnarray}
    \rho - p &=& \frac{1-\gamma}{1+\gamma} f(v),
    \label{14a} \\
    \rho + p &=& f(v),
    \label{14b} \\
    \rho + 3 p &=& \frac{1+3\gamma}{1+\gamma} f(v),
    \label{14c}
\end{eqnarray}
and we require each to be positive. Finally adiabatic stability is satisfied as
\begin{equation}
    \left(\frac{\rho + p}{p} \right) \frac{dp}{d\rho} = \frac{3}{2} > \frac{4}{3}.
\end{equation}

\subsection{Quadratic equation of state}
\label{subsec:quadratic}

Next, we consider the polytropic index $n=1$ or $\Gamma=2$ which gives $p = \gamma\rho^2$. Beginning again with the right-hand side of~(\ref{11}) one finds quadratic equations for density and pressure. Solving these gives
\begin{equation}
    \rho = \frac{1}{2\gamma}\sqrt{1+4\gamma f(v)}-\frac{1}{2\gamma}, \qquad
    p = \left[\frac{1}{2\gamma}\sqrt{1+4\gamma f(v)}-\frac{1}{2\gamma}\right]^2,
    \label{rhoandp}
\end{equation}
where the other root is neglected as it would result in negative $\rho$. The pressure and energy density both vanish when $f(v)=0$. Differentiating the pressure equation yields the sound speed component as
\begin{equation}
    \frac{dp}{d\rho}= 2\rho = \frac{1}{\gamma}\sqrt{1+4\gamma f(v)}-\frac{1}{\gamma}.
    \label{16}
\end{equation}
The energy conditions are given by
\begin{eqnarray}
    \rho - p &=& \frac{1}{\gamma} \sqrt{1+4\gamma f(v)}-\frac{1}{\gamma}-f(v),  \label{17a}\\
    \rho + p &=& f(v), \label{17b}\\
    \rho + 3p &=& 3f(v)+\frac{1}{\gamma}-\frac{1}{\gamma}\sqrt{1+4\gamma f(v)}, \label{17c}
\end{eqnarray}
respectively. The condition $\rho-p>0$ is equivalent to $f(v)<2/\gamma$ which in combination with $f(v)>0$ yields the neat inequality $0<f(v)<2/\gamma$. The dominant energy condition does not introduce any further conditions. 

Then the stability index for the quadratic model takes the form 
\begin{equation}
    \left(\frac{\rho + p}{p} \right) \frac{dp}{d\rho} = 
    \left(\frac{\rho+\gamma\rho^2}{\gamma\rho^2} \right) 2\gamma\rho = 2(1+\gamma\rho) =
    \sqrt{1+4\gamma f(v)}+1,
    \label{18}
\end{equation}
and the requirement of this being greater than $4/3$ also does not introduce any additional restrictions. We are now ready to investigate the physical properties of this solution.

\subsection{Physical analysis of the quadratic model}

\begin{figure}[!htb]
    \centering
    \includegraphics[width=0.60\textwidth]{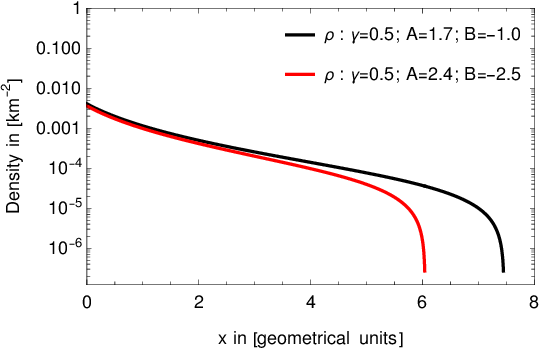}
    \caption{Log plot plot of the energy density versus radial parameter $x=C r^2$. Quadratic equation of state, numerical values $\gamma=0.5$, $A=1.7$, $B=-1.0$ (black), and $\gamma=0.5$, $A=2.4$, $B=-2.5$ (red).}
    \label{fig1}
\end{figure}

In order to examine the model's properties as a candidate representing a physically reasonable compact star, a suitable parameter space must be determined. In the case of the Finch-Skea polytropic model the parameter values $A=2.4$ and $B=-1$ or $A=2.4$ and $b=2.5$ give realistic physical behaviour. We show the typical form of the energy density in Fig.~\ref{fig1}, with the pressure behaving similarly. 
The specifically chosen values for our two constants $A$ and $B$ will become clear below. Both, the energy density and pressure are finite at $x=0$ and vanish for some finite radius, this radius is typically taken to be the radius of the object. Note that we used a logarithmic plot in Fig.~\ref{fig1} to emphasise the radius where the energy density vanishes. Since the speed of sound is proportional to the energy density, this is also well behaved throughout the object. Moreover, as the energy conditions follow largely the energy density it is clear that these are all satisfied, including the adiabatic stability index of Chandrasekhar, as derived in Eq.~(\ref{18}), which again closely follows the energy density.

\subsection{Matching with the exterior of the quadratic model}

In order to complete the study of this model, we will now relate constants $A$ and $B$ to the mass $M$ and radius $R$ of the star by the continuity of the first and second fundamental forms. Alternatively, the vanishing pressure condition is equivalent to the matching of the second fundamental forms through the Israel-Darmois junction condition. 

Let $R$ be the radius where the pressure vanishes. We refer to this as the vanishing pressure surface and use this surface to define the surface of the astrophysical object we are studying. Hypothetically the energy density and pressure could vanish for different radii, however, for polytropes this radius is indeed unique. 

Setting $V=1+R^2$ the vanishing pressure equation gives
\begin{equation}
    \tan V \left(B V^2-A V \left(V^2+1\right)\right) +
    A V^2+B \left(V^2+1\right) V = 0. 
    \label{19}
\end{equation}
We note in passing that one cannot find $V$ (or $R$) explicitly in terms of the two constants $A$ and $B$. Note that the constant $C$ is an arbitrary scaling constant for the radial variable. The matching of the $g_{00}$ metric component yields
\begin{equation}
    \sqrt{\frac{2}{\pi }} (\sin V (A-B V)-\cos V (A V+B))=
    \sqrt{1-\frac{2 M}{R}},
    \label{20}
\end{equation}
where we have set $\Lambda=0$ to give the standard Schwarzschild exterior metric. Simultaneously solving these equations give
\begin{eqnarray} 
    A = A(M,R) &=& -\sqrt{\frac{\pi }{2}}\frac{1}{V^3} \cos V \left(V^2+V \tan V+1\right) 
    \sqrt{1-\frac{2 M}{R}},
    \label{21a} \\[1ex]
    B = B(M,R) &=& \sqrt{\frac{\pi}{2}}\frac{1}{V^3} 
    \left(V \cos V-\left(V^2+1\right) \sin V\right) 
    \sqrt{1-\frac{2 M}{R}}, 
    \label{21b}
\end{eqnarray}
where we now emphasise the $A$ and $B$ are determined by choosing the mass and radius. These constant could now be substituted back into Eq.~(\ref{11}) to find the function $f(v)$ explicitly for different $M$ and $R$. However, at this point it is not clear whether any such choice is in fact compatible with the various physical requirement like $0<f(v)<2$ for all $v$.

Finally, we show some physical characteristics as well as possible values of the parameters $\gamma$, $A$, $B$, and $C$ of five compact stars using our solution in Tables~\ref{table1}. These values are not unique as $\gamma$ was chosen. Using the given values for mass or radius we determine that the previous parameter choices provide a realistic picture.
\begin{table}[!thb]
\centering
\begin{tabular}{||c|cccc||}
\hline
Compact star models &$M$~($M_{\odot}$)& $R$~($km$) & $A$ & $B$  \\
\hline
PSR J0030+0451 (Miller \emph{et al.}~\cite{Miller:2019cac}) & $1.34^{+0.15}_{-0.16}$ & $12.71^{+1.14}_{-1.19}$ & $1.999$& $-1.188$  \\
PSR J0437-4715 (Gonzalez-Caniulef \emph{et al.}~\cite{Gonzalez-Caniulef:2019wzi}) & $1.44^{+0.7}_{-0.07}$ & $13.6^{+0.9}_{-0.8}$ & $1.997$ & $-1.187$  \\ 
Cen X-3 (Rawls \emph{et al.}~\cite{Rawls:2011jw}) & $1.49^{+0.08}_{-0.08}$ & $9.178^{+0.13}_{-0.13}$ & $1.738$ & $-1.033$  \\ 
PSR J1614-2230 (Demorest \emph{et al.}~\cite{Demorest:2010bx}) & $1.97^{+0.04}_{-0.04}$ & $13^{+2}_{-2}$ & $1.790$ & $-1.064$ \\
PSR J0740+6620 (Cromartie \emph{et al.}~\cite{NANOGrav:2019jur}) & $2.14^{+0.20}_{-0.17}$ & $13.7^{+2.6}_{-1.5}$ & $1.768$& $-1.051$ \\
\hline
\end{tabular}
\caption{The values of model constants using mass and radius of different compact stars for the Finch-Skea seed metric via quadratic EoS. We set $\gamma=1/2$ and $C=10^{-3}$.}
\label{table1}
\end{table}
Table~\ref{table2} shows the physical quantities $\rho(0)$, $\rho(R)$, $p(0)$ and $p(0)/\rho(0)$. These were computed using the parameters $A$ and $B$ as given by the mass and radius of the data, again $\gamma=1/2$. We note that these are increasing with increasing mass. Consequently, the radius decreases, which shows that stars become more compact. Note we use the quantity $p(0)/\rho(0)$ as a proxy for a compactness parameter. Other possibilities are the quantity $M/R$ or the average density $M/(4\pi R^3/3)$ which contain similar information. Overall the Finch-Skea metric provides a good working model for astrophysical objects.
\begin{table}[!htb]
\centering
\begin{tabular}{||c|cccc||}
\hline
Compact star models &  $\rho(0)$ & $\rho(R)$ & $p(0)$ & ${p(0)/\rho(0)}$ \\
\mbox{} &  $(g/cm^{3})$ & $(g/cm^{3})$ & $(dyne/cm^{3})$ & $\times 10^{-3}$ \\ 
\hline
PSR J0030+0451 (Miller \emph{et al.}~\cite{Miller:2019cac})  & $6.27\times 10^{15}$ & $4.65\times 10^{15}$& $1.46\times 10^{34}$ & $2.33$ \\
PSR J0437-4715 (Gonzalez-Caniulef \emph{et al.}~\cite{Gonzalez-Caniulef:2019wzi})  & $6.54\times 10^{15}$ & $4.60\times 10^{15}$ & $1.59\times 10^{34}$ & $2.43$ \\ 
Cen X-3 (Rawls \emph{et al.}~\cite{Rawls:2011jw}) & $6.81\times 10^{15}$ & $5.72\times 10^{15}$& $1.72\times 10^{34}$ & $2.53$ \\ 
PSR J1614-2230 (Demorest \emph{et al.}~\cite{Demorest:2010bx}) & $7.08\times 10^{15}$ & $4.99\times 10^{15}$& $1.86\times 10^{34}$ & $2.63$ \\
PSR J0740+6620 (Cromartie \emph{et al.}~\cite{NANOGrav:2019jur})  & $7.36\times 10^{15}$  &$4.94\times 10^{15}$& $2.01\times 10^{34}$& $2.73$ \\
\hline
\end{tabular}
\caption{Physical parameters of the observed stellar toy models for the numerical values of the constant parameters $\gamma$, $A$, $B$ and $C$, as given in Table~\ref{table1} for the Finch-Skea seed metric with quadratic EoS.}
\label{table2}
\end{table}

\section{Vaidya-Tikekar superdense star}

Let us now turn to another well-known solution called Vaidya-Tikekar~\cite{Vaidya:1982zz} with metric ansatz given by
\begin{eqnarray}
    Z&=&\frac{a x+1}{2 a x+1},\\
    y&=&A \sqrt{a x+1}-\frac{2 B}{a}  \left(\sqrt{2 a x+1}-\sqrt{2} \sqrt{a x+1} \tanh^{-1} \left(\sqrt{\frac{2 a x+1}{2 a x+2}}\right)\right),
\end{eqnarray}
where $a$ is a real parameter and $A$ and $B$ are two further constants. This spheroidal spacetime has been shown  to characterise superdense stars with densities of the order $10^{14}$ $g/cm^3$ in Einstein gravity~\cite{Vaidya:1982zz}.

As before, we begin with Eq.~(\ref{11}) using the given metric functions which yields
\begin{equation}
    \rho + p = \frac{C\, a}{1+2ax}\Bigl[
    \frac{2}{1+2ax} + 2 - \frac{4B\sqrt{1+2ax}}{\sqrt{1+ax}}
    \Bigl(2B\frac{\sqrt{1+2ax}}{\sqrt{1+ax}}-aA-2\sqrt{2}B\tanh^{-1}\sqrt{\frac{2 a x+1}{2 a x+2}}\Bigr)^{-1}
    \Bigr] := W(x).
    \label{eqW}
\end{equation}
As in the previous case, we will again consider the quadratic EoS $p=\gamma \rho^2$. This means we can use Eqs.~(\ref{rhoandp})--(\ref{18}) with $W(x)$ used instead of $f(v)$. This means we have, again, the neat inequalities $0<W(x)<2/\gamma$ for a physically reasonable solution. Next, we will determine the constants $A$ and $B$ via the matching with the exterior solutions and then discuss the the physical properties of the solution.

The radial coordinate of the vanishing pressure surface defines the radius of the star. Together with the of the $g_{00}$ components of the metric we find the following relations 
\begin{align}
    A &= A(M,R) = \frac{\sqrt{1-\frac{2M}{R}}}{(1+2a C R^2)} 
    \Bigl(
    -\frac{1}{\sqrt{1+a C R^2}} + 
    2\sqrt{2}\frac{1+a C R^2}{\sqrt{1+2a C R^2}} \tanh^{-1}\sqrt{\frac{1+ 2 a C R^2}{2+2 a C R^2}}
    \Big), 
    \label{50ba} \\[1ex]
    B &= B(M,R) = -a\frac{1+a C R^2}{(1+2a C R^2)^{3/2}}
    \sqrt{1-\frac{2M}{R}}. 
    \label{50b}
\end{align}
Note that the constants $a$ and $C$ may be chosen arbitrarily. For simplicity we choose $a=C=1$ which means we are left with the two constants $A$ and $B$. These are uniquely related to the mass and the radius of the object.

\begin{figure}[!htb]
    \centering
    \includegraphics[width=0.60\textwidth]{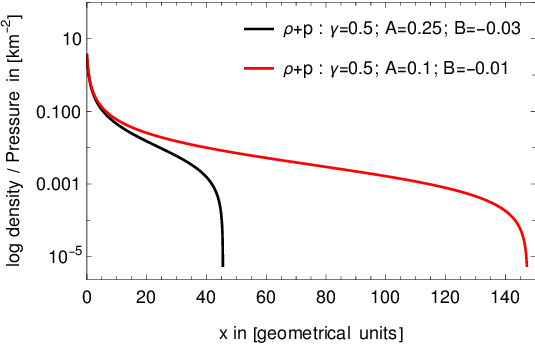}
    \caption{Log plot of energy density plus pressure (the function $W$, Eq.~(\ref{eqW})) versus radial parameter $x$. We choose the numerical values $A=0.25$, $B=-0.03$ (black) and $A=0.1$, $B=-0.01$ (red). We note that the pressure vanishes for finite radius.}
    \label{new2}
\end{figure}

Physically plausible models are found for the following values: $A=0.25$, $B=-0.03$ or $A=0.1$, $B=-0.01$, as shown in Figure~\ref{new2}. As before, we choose a logarithmic plot to emphasise the key features of this solutions. This shows the function $W=\rho+p$, Eq.~(\ref{eqW}). We note that vanishing pressure surface's location is very sensitive to variation of the parameters while the central pressure is less affected.

It is straightforward to verify that the sound speed index satisfies $0\le dp/d\rho \le 1$ as required for causality. The three energy energy conditions are all satisfied. This is largely expected from the general discussion shown in Sec.~\ref{subsec:quadratic}.

Finally, we compare this model with five compact star candidates and we derive some physical quantities, including $\rho(0)$, $\rho(R)$, $p(0)$ and $p(0)/\rho(0)$, where the latter quantity is used as a compactness measure. We show possible numerical values of the parameters $\gamma$, $A$, $B$, $C$ and $a$ for the quadratic model we are considering. These are shown in Tables~\ref{table3} and~\ref{table4}. These resulting quantities allow us to model a mass range from $1.18~M_{\odot}$ to $2.34~M_{\odot}$.

\begin{table}[!thb]
\centering
\begin{tabular}{||c|cccc||}
\hline
Compact star models &$M$~($M_{\odot}$)& $R$~($km$) & $A$ & $B$  \\
\hline
PSR J0030+0451 (Miller \emph{et al.}~\cite{Miller:2019cac}) & $1.34^{+0.15}_{-0.16}$ & $12.71^{+1.14}_{-1.19}$ & $0.234$ & $-0.0231$  \\
PSR J0437-4715 (Gonzalez-Caniulef \emph{et al.}~\cite{Gonzalez-Caniulef:2019wzi}) & $1.44^{+0.7}_{-0.07}$ & $13.6^{+0.9}_{-0.8}$ & $0.238$ & $-0.0231$  \\ 
Cen X-3 (Rawls \emph{et al.}~\cite{Rawls:2011jw}) & $1.49^{+0.08}_{-0.08}$ & $9.178^{+0.13}_{-0.13}$ &  $0.223$ & $-0.0215$  \\ 
PSR J1614-2230 (Demorest \emph{et al.}~\cite{Demorest:2010bx}) & $1.97^{+0.04}_{-0.04}$ & $13^{+2}_{-2}$ & $0.206$ & $-0.0202$ \\
PSR J0740+6620 (Cromartie \emph{et al.}~\cite{NANOGrav:2019jur}) & $2.14^{+0.20}_{-0.17}$ & $13.7^{+2.6}_{-1.5}$ & $0.196$ & $-0.0189$ \\
\hline
\end{tabular}
\caption{The values of model constants using mass and radius of different compact stars for the Vaidya-Tikekar seed metric with quadratic EoS. We set $\gamma=1/2$, $a=5\times 10^{-3}$ and $C=1$.}
\label{table3} 
\end{table}
\begin{table}[!htb]
\centering
\begin{tabular}{||c|cccc||}
\hline
Compact star models & $\rho(0)$ & $\rho(R)$ & $p(0)$ & $p(0)/\rho(0)$ \\
\mbox{} &   $(g/cm^{3})$ & $(g/cm^{3})$ & $(dyne/cm^{3})$ &  $\times 10^{-3}$\\ \hline
PSR J0030+0451 (Miller \emph{et al.}~\cite{Miller:2019cac}) &  $8.18897\times 10^{15}$ & $6.38189\times 10^{15}$& $2.48996\times 10^{34}$ & $3.04062$\\
PSR J0437-4715 (Gonzalez-Caniulef \emph{et al.}~\cite{Gonzalez-Caniulef:2019wzi}) &  $8.19005\times 10^{15}$ & $6.17606\times 10^{15}$ & $2.49061\times 10^{34}$ & $3.04102$ \\ 
Cen X-3 (Rawls \emph{et al.}~\cite{Rawls:2011jw}) &  $8.18991\times 10^{15}$ & $7.15423\times 10^{15}$& $2.49053\times 10^{34}$ & $3.04097$ \\ 
PSR J1614-2230 (Demorest \emph{et al.}~\cite{Demorest:2010bx}) &  $8.18933\times 10^{15}$ & $6.31512\times 10^{15}$ & $2.49018\times 10^{34}$ & $3.04076$ \\
PSR J0740+6620 (Cromartie \emph{et al.}~\cite{NANOGrav:2019jur}) &  $8.19017\times 10^{15}$ & $6.15279\times 10^{15}$ & $2.49068\times 10^{34}$ & $3.04107$ \\
\hline
\end{tabular}
\caption{Physical parameters of the observed stellar models for the given numerical values of Table~\ref{table3}.}
\label{table4}
\end{table}

\section{Conclusion}
\label{Sec6}

Finding exact solutions for perfect fluid stars with polytropic equations of state is difficult in the context of relativistic astrophysics. Even the simplest equation of state, namely a linear equation, does not give physically meaningful solutions modelling a compact object. While a polytropic equation of state would be a more realistic model for a compact object, the resulting differential equations have no known exact solutions in General Relativity.

In this paper we circumvent this problem by approaching the field equations differently and by working with the trace-free version of the Einstein field equations which contain more freedom in the choice of functions. Our proposal is to generate exact models of polytropic stars in this unimodular model. One can motivate this approach by accepting that stars loose energy through radiation which is not covered by the standard Einstein field equations. This allows us to select one part of the metric function of a known exact stellar model as an input for the field equations. The energy density and pressure are intrinsically joined in the trace-free field equations, one always encounters the term $\rho+p$. Consequently, choosing an equation of state will allow one to separate these two quantities and find appropriate solutions. For polytropic equations of state with small polytropic index one can generally solve these equations explicitly. 

To illustrate the approach, we considered the Finch-Skea seed metric and then imposed a polytropic equation of state of the form $p=\gamma \rho^2$. This quadratic model is easy to deal with analytically but gives nonetheless a realistic stellar model. A suitable parameter space was found such that elementary physical requirements were met which included the existence of a vanishing pressure surface, subluminal sound speeds, satisfies energy conditions, and adiabatic stability according to Chandrasekhar. The resulting solution was regular with finite energy density and pressure through the star. We also considered the more general Vaidya-Tikekar superdense star model, again assuming a quadratic equation of state.

\end{document}